\documentclass{article}

\usepackage{amssymb}

\newcommand{\be}{\begin{equation}}
\newcommand{\ee}{\end{equation}}
\newcommand{\lb}{\label}
\newcommand{\ol}{\overline}

\newcommand{\ba}{{\bf a}}
\newcommand{\bg}{{\bf g}}
\newcommand{\bu}{{\bf u}}
\newcommand{\bv}{{\bf v}}
\newcommand{\bw}{{\bf w}}
\newcommand{\bx}{{\bf x}}
\newcommand{\br}{{\bf r}}

\newcommand{\BF}{{\bf F}}

\newcommand{\bW}{{\bf W}}

\newcommand{\barphi}{{\mbox{\boldmath $\varphi$}}}

\newcommand{\grad}{{\mbox{\boldmath $\nabla$}}}
\newcommand{\bdot}{{\mbox{\boldmath $\cdot$}}}

\newcommand{\bzed}{{\mbox{\boldmath $0$}}}

\newcommand{\scirc}{{\scriptstyle \circ}}

\newtheorem{Prop}{Proposition}

\begin{document}
\title{Stochastic Least-Action Principle for the \\ Incompressible
Navier-Stokes Equation}
\author{Gregory L. Eyink\\
{\it Department of Applied Mathematics \& Statistics}\\
{\it The Johns Hopkins University}\\
{\it 3400 N. Charles Street}\\
{\it Baltimore, MD 21218}}
\date{ }
\maketitle
\begin{abstract}
We formulate a stochastic least-action principle for solutions of the
incompressible
Navier-Stokes equation, which formally reduces to Hamilton's principle for the
incompressible Euler solutions in the case of zero viscosity. We use this
principle
to give a new derivation of a stochastic Kelvin Theorem for the Navier-Stokes
equation,
recently established by Constantin and Iyer, which shows that this stochastic
conservation
law arises from particle-relabelling symmetry of the action. We discuss  issues
of
irreversibility, energy dissipation, and the inviscid limit of Navier-Stokes
solutions
in the framework of the stochastic variational principle. In particular, we
discuss the
connection of the stochastic Kelvin Theorem with our previous ``martingale
hypothesis''
for fluid circulations in turbulent solutions of the incompressible Euler
equations.
\end{abstract}

\section{Introduction}

Alternative formulations of standard equations can be very illuminating and can
cast new light on old problems. As just one example, consider how Feynman's
path-integral solution of the Schr\"{o}dinger equation enabled intuitive new
approaches
to difficult problems with many-degrees-of-freedom, such as quantum
electrodynamics
and superfluid helium. In this same spirit, many different mathematical
formulations
have been developed for the equations of classical hydrodynamics, both ideal
and
non-ideal. Recently, Constantin and Iyer \cite{ConstantinIyer08} have presented
a
very interesting representation of solutions of the incompressible
Navier-Stokes
equation by averaging over stochastic Lagrangian trajectories in the Weber
formula
\cite{Weber1868} for incompressible Euler solutions. Their formulation is a
nontrivial application of the method of stochastic  characteristics, well-known
in pure mathematics \cite{Kunita90} (Chapter 6), in theoretical physics
\cite{ShraimanSiggia94,Falkovichetal01} and in engineering modeling
\cite{Sawford01,EgbertBaker84}.   The characterization of the Navier-Stokes
solutions
in \cite{ConstantinIyer08}  is through a nonlinear fixed-point problem, since
the velocity
field that results from the average over stochastic trajectories must be the
same as that
which advects the fluid particles. Constantin and Iyer have shown that their
stochastic representation implies remarkable properties of Navier-Stokes
solutions in close analogy to those of ideal Euler solutions, such as a
stochastic Kelvin
Theorem for fluid circulations and a stochastic Cauchy formula for the
vorticity field.

In this paper, we point out some further remarkable features of the stochastic
Lagrangian
formulation of \cite{ConstantinIyer08}. Most importantly, we show that the
nonlinear
fixed-point problem that characterizes the Navier-Stokes solution is, in fact,
a variational
problem which generalizes the well-known Hamilton-Maupertuis least-action
principle
for incompressible Euler solutions \cite{Salmon88}. We shall demonstrate this
result by
a formally exact calculation, at the level of rigor of theoretical physics. A
more careful
mathematical proof, with set-up of relevant function spaces, precise
definitions of
variational derivatives, etc. shall be given elsewhere. Closely related
stochastic
variational formulations of incompressible Navier-Stokes solutions have been
developed recently by others \cite{Rapoport00,Gomes05,CiprianoCruzeiro07}
and a detailed comparison with these approaches will also be made in future
work.

Our variational formulation sheds some new light on a basic proposition of
\cite{ConstantinIyer08}, the stochastic Kelvin Theorem which was established
there
for smooth Navier-Stokes solutions at any finite Reynolds number. We show that
this result is a consequence of particle-relabelling symmetry of our stochastic
action
functional for Navier-Stokes solutions, in the same manner as  the usual Kelvin
Theorem arises from particle-relabelling symmetry of the standard action
functional
for Euler solutions \cite{Salmon88}. This result strengthens the conjecture
made
by us in earlier work \cite{Eyink06,Eyink07} that a ``martingale property'' of
circulations
should hold for generalized solutions of the incompressible Euler equations
obtained
in the zero-viscosity limit.
Indeed, the stochastic variational principle for Navier-Stokes solutions
considered in
the present work is very closely similar to a stochastic least-action principle
for generalized
solutions of incompressible Euler equations that was developed by Brenier
\cite{Brenier89,Brenier93,Brenier03}. One of the arguments advanced for
the ``martingale property'' in \cite{Eyink06} was particle-relabelling symmetry
in a Brenier-type variational formulation of generalized Euler solutions.
That argument, however, did not distinguish an arrow of time,  so that fluid
circulations might satisfy the martingale property either forward or backward
in time. It was subsequently argued in \cite{Eyink07} that the
backward-martingale
property is the correct one, consistent with time-irreversibility in the limit
of vanishing viscosity. The present work shows that a small but positive
viscosity
indeed selects the backward martingale  property, as expected for a causal
solution.

\section{The Action Principle}

The action principle formulated here for Navier-Stokes solutions involves {\it
stochastic flows}
\cite{Kunita90}. The relevant flows are those which solve a {\it backward Ito
equation}:
\be \left\{\begin{array}{l}

\hat{d}_t\bx^\varpi(\ba,t)=\bu^\varpi(\bx^\varpi(\ba,t),t)dt
                            +\sqrt{2\nu}\hat{d}\bW^\varpi(t),t<t_f \cr
                            \bx^\varpi(\ba,t_f)=\ba.  \cr
                      \end{array} \right. \lb{sde} \ee
Here $\bW^\varpi(t),\,\,t\in[t_0,t_f]$ is a $d$-dimensional Brownian motion on
a probability
space $(\Omega,P,{\cal F})$ which is adapted to a two-parameter filtration
${\cal F}_t^{t'}$
of sub-$\sigma$-fields of ${\cal F},$ with $t_0\leq t<t'\leq t_f.$ Thus,
$\bW^\varpi(s)-\bW^\varpi(s')$ is ${\cal F}_t^{t'}$-measurable for all
$t\leq s<s'\leq t'.$  The constant $\nu$ that appears in the amplitude of the
white-noise
term in the SDE (\ref{sde}) will turn out to be the kinematic viscosity in the
Navier-Stokes
equation. Note that, for such an additive noise as appears in (1), the
(backward) Ito
and Stratonovich equations are equivalent.

In order to describe the space of flow maps which appear in the action
principle, we
must  make a few slightly technical, preliminary remarks. The random velocity
field
$\bu^\varpi(\br,t)$ in equation (\ref{sde}) is assumed to be smooth and, in
particular,
continuous in time, as well as adapted to the filtration ${\cal F}_t^{t_f},
\,\,\,t<t_f$ backward in time. It then follows from standard theorems (e.g. see
Corollary 4.6.6 of \cite{Kunita90}) that the solution $\bx^\varpi(\ba,t)$ of
(\ref{sde})
is a backward semi-martingale of flows of diffeomorphisms. Conversely, any
backward semi-martingale of flows of diffeomorphisms has a backward
Stratonovich
random infinitesimal generator $\mathring{\BF}^\varpi(\br,t)$ which is a
spatially-smooth
backward semi-martingale (e.g. see Theorem 4.4.4 of \cite{Kunita90}). The class
of
such flows for which the martingale part of the generator is
$\sqrt{2\nu}\bW^\varpi(t)$
and for which the bounded-variation part of the generator is
absolutely-continuous with
respect to $dt$ coincides with the class of solutions of equations of form (1),
for
all possible choices of $\bu^\varpi(\br,t)$ .  Clearly,  the random fields
$\bu^\varpi(\br,t)$
and $\bx^\varpi(\ba,t)$ uniquely determine each other. We consider here the
incompressible case, where $\bu^\varpi(\br,t)$ is divergence-free and
$\bx^\varpi(\ba,t)$
is volume-preserving a.s.

The action is defined as a functional of the backward-adapted random velocity
fields
$\bu^\varpi(\br,t)$---or, equivalently, of the random flow maps
$\bx^\varpi(\ba,t)$---by
the formula
\be S[\bx] =\int P(d\varpi) \int_{t_0}^{t_f} dt\int
d^dr\,\frac{1}{2}|\bu^\varpi(\br,t)|^2
      \lb{action} \ee
when this is well-defined and as $+\infty$ otherwise. The {\it variational
problem}
(VP) is to find a stationary point of this action such that
$\bx^\varpi(\ba,t_f)=\ba$ and
$\bx^\varpi(\ba,t_0)=\barphi^\varpi(\ba)$ for $P-a.e.\,\,\,\varpi,$ where
$\barphi^\varpi(\ba)$
is a given random field of volume-preserving diffeomorphisms of the flow
domain.  It is
interesting that this problem is very similar to that considered by Brenier
\cite{Brenier89,Brenier93,Brenier03} for generalized Euler solutions. The above
problem
leads instead to the incompressible Navier-Stokes equation, in the following
precise sense:

\begin{Prop}.  A stochastic flow $\bx^\varpi(\ba,t)$ which satisfies both the
initial and final
conditions is a solution of the above variational problem if and only if
$\bu^\varpi(\br,t)$
solves the incompressible Navier-Stokes equation with viscosity $\nu>0$
\be \partial_t\bu^\varpi+(\bu^\varpi\bdot\grad)\bu^\varpi=-\grad p^\varpi
+\nu\bigtriangleup\bu^\varpi,\,\,\,\,\,\,\,\,\,\,P-a.s. \lb{NS} \ee
where kinematic pressure $p^\varpi$ is chosen so that $\grad\bdot\bu^\varpi=0.$
\end{Prop}

\noindent {\it Proof:} Making a variation $\delta\bu^\varpi(\br,t)$ in the
random velocity
field, the equation (\ref{sde}) becomes
\be \left\{\begin{array}{l}
     \hat{d}_t\delta\bx^\varpi(\ba,t) =
[\delta\bx^\varpi(\ba,t)\bdot\grad_r\bu^\varpi(\bx^\varpi,t)
+\delta\bu^\varpi(\bx^\varpi,t)]dt,
     \,\,\,\,\,t<t_f  \cr
     \delta\bx^\varpi(\ba,t_f)=\bzed.  \cr
                      \end{array} \right. \lb{vareq} \ee
Since the VP requires that $\bx^\varpi(\ba,t_0)=\barphi^\varpi(\ba),$ one can
only consider
variations such that, also, $\delta\bx^\varpi(\ba,t_0)=\bzed.$ (We shall
consider below an
alternative approach with a Lagrange multiplier that permits unconstrained
variations.)
This equation may also be written as
\be
\hat{d}_t\delta\bx^\varpi(\ba,t)
-(\grad_r\bu^\varpi(\bx^\varpi,t))^\top\delta\bx^\varpi(\ba,t)dt
=\delta\bu^\varpi(\bx^\varpi,t)dt \lb{vareq-alt}
\ee
for $t<t_f$ and then easily solved by Duhamel's formula (backward in time)
to give  $\delta\bx^\varpi(\ba,t)$ in terms of $\delta\bu^\varpi(\br,t).$ Since
the martingale
term vanished under variation, the process  $\delta\bx^\varpi(\ba,t)$ is of
bounded
variation and clearly adapted to the backward filtration ${\cal
F}_t^{t_f},\,\,t<t_f.$
Conversely, any such flow variation will determine the corresponding velocity
variation
$\delta\bu^\varpi(\br,t)$ by the equation (\ref{vareq-alt}) directly. Lastly,
note that the
volume-preserving condition $\det(\grad_a\bx^\varpi(\ba,t))=1$ becomes
\be \grad_r\bdot \delta\bx^\varpi(\ba^\varpi,t)=0 \lb{del-inc} \ee
under variation, where $\ba^\varpi(\br,t)$ is the ``back-to-labels'' map
inverse to the
flow map $\bx^\varpi(\ba,t)$. Because these maps are diffeomorphisms, we see
that the Eulerian variation of the flow map,
$\delta\bar{\bx}^\varpi(\br,t)\equiv
\delta\bx^\varpi(\ba^\varpi(\br,t),t),$ is an arbitrary divergence-free field.

With these preparations, we obtain for the variation of the action
(\ref{action}):
\begin{eqnarray}
\delta S[\bx] & = & \int P(d\varpi) \int_{t_0}^{t_f} dt\int
d^dr\,\,\bu^\varpi(\br,t)\bdot \delta\bu^\varpi(\br,t) \cr
 \, & = & \int P(d\varpi) \int d^da\int_{t_0}^{t_f}
\,\,\bu^\varpi(\bx^\varpi(\ba,t),t)\bdot
 \left[
\hat{d}_t\delta\bx^\varpi(\ba,t)-\delta\bx^\varpi(\ba,t)\bdot
\grad_r\bu^\varpi(\bx^\varpi,t)dt \right] \cr
\, & = & -\int P(d\varpi) \int d^da  \int_{t_0}^{t_f}\,\,\left[\hat{d}_t
\bu^\varpi(\bx^\varpi,t)
+\grad_r\left.\left(\frac{1}{2}
|\bu^\varpi|^2\right)\right|_{\bx^\varpi}dt\right]\bdot \delta\bx^\varpi(\ba,t)
\lb{var-act} \end{eqnarray}
In the second line we employed (\ref{vareq-alt}).  In the third  line we
integrated by parts, using
the facts that  $\delta\bx^\varpi(\ba,t_f)=\delta\bx^\varpi(\ba,t_0)=0$ and
that $\delta\bx^\varpi(\ba,t)$
is a bounded variation process, so that the quadratic variation vanishes:
$\hat{d}_t\langle
\bu^\varpi(\bx^\varpi,t),\delta\bx^\varpi(\ba,t)\rangle=0.$ We note that the
final
gradient term vanishes, because $\delta\bar{\bx}^\varpi(\br,t)$ is
divergence-free. We can
evaluate the remaining term using the chain rule
\begin{eqnarray}
\hat{d}_t\bu^\varpi(\bx^\varpi,t)=\partial_t\bu^\varpi(\bx^\varpi,t)dt
          + (\bx^\varpi(\ba,\scirc\hat{d}t)\bdot\grad)\bu^\varpi(\bx^\varpi,t),
\lb{DuDt-S} \end{eqnarray}
in terms of the backward Stratonovich differential. This result can also be
written using Ito calculus.
Calculating from (\ref{sde}) and (\ref{DuDt-S}) the quadratic variation
$$ \sqrt{2\nu}\hat{d}_t\langle
W_j^\varpi(t),\partial_{x_j}\bu^\varpi(\bx^\varpi,t)\rangle
               = 2\nu \bigtriangleup\bu^\varpi(\bx^\varpi,t)dt, $$
one obtains the backward Ito equation
\be
\hat{d}_t\bu^\varpi(\bx^\varpi,t)=
[\partial_t\bu^\varpi+(\bu^\varpi\bdot\grad)\bu^\varpi-
      \nu\bigtriangleup\bu^\varpi](\bx^\varpi,t)dt
      + \sqrt{2\nu}(\hat{d}\bW^\varpi(t)\bdot\grad)\bu^\varpi(\bx^\varpi,t),
\lb{DuDt-I}
\ee
A crucial point is that the martingale part of (\ref{DuDt-I}) vanishes when the
expression
is substituted back into (\ref{var-act}), since both
$(\grad\bu^\varpi)(\bx^\varpi,t)$ and
$\delta\bx^\varpi(\ba,t)$ are adapted to the backward filtration ${\cal
F}_t^{t_f},\,\,t<t_f.$
Thus, the final result is
\be
\delta S[\bx] = \int P(d\varpi) \int_{t_0}^{t_f} dt\int d^dr
        \left[\partial_t\bu^\varpi+(\bu^\varpi\bdot\grad)\bu^\varpi
        -\nu\bigtriangleup\bu^\varpi\right](\br,t)\bdot
\delta\bar{\bx}^\varpi(\br,t).
\lb{var-fin} \ee
Since the integrands are smooth in space and continuous in time and since the
flow variation is an arbitrary divergence-free field, the theorem statement
follows. $\Box$

\vspace{.2in}

\noindent There are alternative formulations of the VP which should be
mentioned. Rather
than performing the variation with the constraint
$\bx^\varpi(\ba,t_0)=\barphi^\varpi(\ba),$
one can instead modify the action with a Lagrange multiplier term:
\be S'[\bx,\bv_0]=S[\bx] + \int P(d\varpi) \int d^da\,\,\bv^\varpi_0(\ba)\bdot
[\bx^\varpi(\ba,t_0)-\barphi^\varpi(\ba)].
\lb{mod-act} \ee
Varying with respect to the Lagrange multiplier $\bv_0^\varpi$ yields the
constraint, whereas
an unconstrained variation with respect to $\bx^\varpi$ yields
\begin{eqnarray}
\delta S'[\bx,\bv_0]  &=&  \int P(d\varpi) \int_{t_0}^{t_f} dt\int d^dr
        \left[\partial_t\bu^\varpi+(\bu^\varpi\bdot\grad)\bu^\varpi
        - \nu\bigtriangleup\bu^\varpi\right](\br,t)\bdot
\delta\bar{\bx}^\varpi(\br,t) \cr
&& \,\,\,\,\,\, +\int P(d\varpi) \int
d^da\,\,[\bv^\varpi_0(\ba)-\bu^\varpi(\bx^\varpi(\ba,t_0),t_0)]
    \bdot\delta \bx^\varpi(\ba,t_0). \lb{del-mod-act} \end{eqnarray}
The second term on the righthand side arises partly from the Lagrange
multiplier
term and partly from integration-by-parts in time. It follows that
$\bv_0^\varpi(\ba)$ can be
identified as the Lagrangian fluid velocity at the initial time $t_0$ and,
likewise,
$\bu_0^\varpi(\br)=\bv_0^\varpi((\barphi^\varpi)^{-1}(\br))$ is the Eulerian
fluid velocity at
time $t_0.$ Another alternative is to add a Lagrange multiplier term $\int
P(d\varpi)
\int_{t_0}^{t_f} dt\int
d^da\,\,\pi^\varpi(\ba,t)\ln\det(\grad_a\bx^\varpi(\ba,t))$ for the
incompressibility constraint and to allow variations over flows which are not
volume-preserving. In that case $\pi^\varpi(\ba,t)$ is the Lagrangian pressure
field and $p^\varpi(\br,t)=\pi^\varpi(\ba^\varpi(\br,t),t)$ is the Eulerian
pressure.

There are several mathematical questions that deserve to be pursued. Some
technical issues remain, e.g. the precise degree of smoothness of solutions
required to make the above argument fully rigorous, etc. It would also be very
interesting
to know under what conditions the solution of the VP corresponds to a minimum
of the action and not just a stationary point. Although we have characterized
the solutions of the VP, we have not proved either their existence or their
uniqueness.
We just remark on the latter point that a unique stationary point certainly
exists if the
initial velocity $\bu_0(\br)$ is deterministic and  if the Navier-Stokes
equation has
a unique solution $\bu(\br,t)$ over the time interval $[t_0,t_f]$ for that
initial datum.
This will be the case, for example, if the initial velocity is smooth enough
and the
Reynolds number $Re=UL/\nu$ is low enough. In that case, the solution of the VP
is also deterministic and is given by the corresponding Navier-Stokes solution.

\section{The Stochastic Kelvin Theorem}

We now mention a closely related result of \cite{ConstantinIyer08}:
\begin{Prop} {\bf (Constantin \& Iyer, 2008)}. The following two properties for
a divergence-free
velocity field  $\bu(\br,t)$ are equivalent:  (i) For all closed, rectifiable
loops $C$ and for any
pair of times $t_0\leq t<t'\leq t_f,$
\be \oint_C \bu(\ba,t')\bdot d\ba=
     \int P(d\varpi)\left[\oint_{\bx^\varpi_{t',t}(C)} \bu(\br,t)\bdot d\br
\right],
      \lb{stoch-kelvin} \ee
where $\bx^\varpi_{t',t}(\ba)$ are the stochastic backward flows which solve
equation (\ref{sde}) with velocity $\bu(\br,t)$ for times $t<t'$ with final
condition
$\bx^\varpi_{t',t'}(\ba)=\ba$;  and, (ii) the velocity $\bu(\br,t)$ satisfies
the  incompressible
Navier-Stokes equation over the time-interval $[t_0,t_f].$
\end{Prop}

\noindent This is just a slight restatement of Theorem 2.2 and Proposition 2.9
of
\cite{ConstantinIyer08}. The result (\ref{stoch-kelvin}) is a stochastic
version of the
Kelvin Theorem on conservation of circulations for incompressible Euler
solutions.
Although circulations are not conserved for Navier-Stokes solutions in the
usual sense,
(\ref{stoch-kelvin}) states that circulations on loops advected by the
stochastic Lagrangian
flow are a martingale backward in time.  This property of the Navier-Stokes
solutions is
closely related to the ``martingale conjecture'' of Eyink \cite{Eyink06} for
generalized
Euler solutions obtained in the limit $\nu\rightarrow 0.$ This connection will
be discussed
in detail in the next section.

It is well-known that the Kelvin Theorem for incompressible Euler equations can
be
derived by the least-action principle as a consequence of an
infinite-dimensional symmetry
\cite{Salmon88}, called ``particle-relabelling symmetry'' and corresponding to
the group of
volume-preserving diffeomorphisms of the flow domain.  This may be done by
applying
the general Noether Theorem relating symmetries and conservation laws. In this
section,
we shall show that the result of Proposition 2 can be similarly derived from
the stochastic
action-principle of section 1 as a consequence of particle-relabelling
symmetry.  We shall
not make use of the Noether Theorem but, following Salmon \cite{Salmon88},
shall instead
employ a more direct method of Lanczos \cite{Lanczos70} based on time-dependent
symmetry
transformations.

Suppose given a smooth 1-parameter family $\{\barphi(\ba,t),
\,\,t\in[t_0,t_f]\}$ of volume-preserving diffeomorphisms satisfying
$\barphi(\ba,t_f)=
\barphi(\ba,t_0)=\ba.$ Then any incompressible flow $\bx(\ba,t)$ may be
deformed
into another such flow
\be \bx_\varphi(\ba,t)\equiv \bx(\barphi(\ba,t),t) \lb{x-deform} \ee
with initial and final values the same. It follows furthermore from
(\ref{x-deform}) by
chain rule that
\be \hat{d}_t\bx_\varphi(\ba,t) =
\hat{d}_t\bx(\bar{\ba},t)+(\dot{\barphi}(\ba,t)
        \bdot\grad_{\bar{a}})\bx(\bar{\ba},t)dt, \lb{x-deform-dot} \ee
for $\bar{\ba}=\barphi(\ba,t).$  If $\bx^\varpi(\ba,t)$ is the solution of the
stochastic
equation (\ref{sde}), then (\ref{x-deform-dot}) implies that
$\bx^\varpi_\phi(\ba,t)$
also satisfies (\ref{sde}) for the modified velocity field
\be  \bu_\phi^\varpi(\br,t)=\bu^\varpi(\br,t)+ (\dot{\barphi}(\ba^\varpi,t)
        \bdot\grad_{\bar{a}})\bx^\varpi(\bar{\ba}^\varpi,t), \lb{u-deform} \ee
where we employ the shorthands $\ba^\varpi=\ba^\varpi(\br,t)$ and
$\bar{\ba}^\varpi=\barphi(\ba^\varpi(\br,t),t).$ It is easy to see from
(\ref{u-deform})
that $\bu_\phi^\varpi(\br,t)$ is adapted to the backward filtration ${\cal
F}_t^{t_f},$
$t<t_f$ whenever the original velocity $\bu^\varpi(\br,t)$ is adapted.

In infinitesimal form, $\barphi(\ba,t)=\ba + \varepsilon
\bg(\ba,t)+O(\epsilon^2)$
with $\grad_a\bdot\bg(\ba,t)=0$ and $\bg(\ba,t_f)=\bg(\ba,t_0)=\bzed.$ The
formula (\ref{u-deform}) then yields the velocity variation
\be \delta\bu^\varpi(\br,t) = \varepsilon
(\dot{\bg}(\ba^\varpi,t)\bdot\grad_{a})
       \bx^\varpi(\ba^\varpi,t)+O(\varepsilon^2). \lb{inf-u-deform} \ee
The corresponding variation of the action is thus [see the first line of
(\ref{var-act})]:
\be 0=\delta S[\bx]=\varepsilon \int P(d\varpi) \int_{t_0}^{t_f} dt\int d^da
\,\,
            \dot{\bg}(\ba,t)\bdot \bw^\varpi(\ba,t) + O(\varepsilon^2),
\lb{inf-S-deform} \ee
where $\bw^\varpi(\ba,t)$ is the {\it stochastic Weber velocity}
\cite{ConstantinIyer08,Weber1868,Salmon88}
\be \bw^\varpi(\ba,t) = \grad_{a}\bx^\varpi(\ba,t)\bdot
\bu^\varpi(\bx^\varpi(\ba,t),t).
       \lb{weber} \ee
We can conclude that the $O(\epsilon)$ variation in (\ref{inf-S-deform}) must
vanish
for any smooth divergence-free function $\bg(\ba,t)$ with
$\bg(\ba,t_f)=\bg(\ba,t_0)=\bzed.$
Taking limits of such functions, one may approximate a divergence-free
distribution
of the form
\be \bg_{C,t,t'}(\ba,\tau) = \chi_{[t',t]}(\tau) \oint_C \delta^d(\ba-\ba')\,
d\ba'
\lb{sing-g} \ee
for any closed, rectifiable loop $C$ and any $t_f>t'>t>t_0.$ Here
$\chi_{[t,t']}(\tau)$
is the characteristic function of the interval $[t,t']$ and
$\delta^d(\ba-\ba')$ is the Dirac
delta-distribution. If we use the property of the Weber velocity that
\be \oint_C \bw^\varpi(\ba',t)\bdot d\ba'
       = \oint_{\bx^\varpi(C,t)}\bu^\varpi(\br,t)\bdot d\br,  \lb{w-circ} \ee
it then follows that
\be \int P(d\varpi) \oint_{\bx^\varpi(C,t')}\bu^\varpi(\br,t')\bdot d\br
= \int P(d\varpi) \oint_{\bx^\varpi(C,t)}\bu^\varpi(\br,t)\bdot d\br,
\lb{circ-ttp} \ee
for any $t_f>t'>t>t_0.$ Taking the limit $t'\rightarrow t_f$ allows us to
identify the
above constant average as $\oint_{C}\ol{\bu}(\br,t_f)\bdot d\br$ with $\ol{\bu}
(\br,t)=\int P(d\varpi)\bu^\varpi(\br,t),$ since
$\bx^\varpi(\ba,t_f)=\ba\,\,\,\,P-a.s.$
When the velocity field that solves the VP is deterministic, i.e. corresponds
to a
unique Navier-Stokes solution $\bu(\br,t)$, then this result gives the
statement
(\ref{stoch-kelvin}) of  Proposition 2 for the special case where $t'=t_f.$
However,
the VP may be applied not only over the entire interval $[t_0,t_f],$ but over
any
subinterval $[t,t']$ as well and this yields the general case.

\section{Irreversibility and the Zero-Viscosity Limit}

At first sight, it is strange to obtain the dissipative Navier-Stokes equation
from
a principle of least-action, which ordinarily leads to time-reversible
equations.
There is no paradox, however, since an ``arrow-of-time'' is built into the
stochastic
action-principle of Section 1. We may say that this is a  {\it causal}
variational
principle, since labels are assigned at the final time and variations are
over prior histories.  The VP may be recast instead to be anti-causal, with
fluid particle labels assigned at  the initial time $t_0$ and with flow maps
solving a forward Ito equation:
\be \left\{\begin{array}{l}
       d_t\bx^\varpi(\ba,t)= \bu^\varpi(\bx^\varpi(\ba,t),t)dt
                            +\sqrt{2\nu}\,d\bW^\varpi(t), \,\,\,\,\,\,\, t>t_0 \cr
                            \bx^\varpi(\ba,t_0)=\ba.  \cr
                      \end{array} \right. \lb{anti-sde} \ee
The random velocity field $\bu^\varpi(\br,t)$ must now be adapted
to the forward filtration ${\cal F}_{t_0}^t,\,\,t>t_0.$  An exact analogue of
Proposition 1
holds, but with the conclusion that the velocity must satisfy
\be \partial_t\bu^\varpi+(\bu^\varpi\bdot\grad)\bu^\varpi=-\grad p^\varpi
-\nu\bigtriangleup\bu^\varpi,\,\,\,\,\,\,\,\,\,\,P-a.s. \lb{anti-NS} \ee
or the {\it negative-viscosity} Navier-Stokes equation. An analogue of
Proposition 2 also holds, in the form
\be \oint_C \bu(\ba,t')\bdot d\ba =
     \int P(d\varpi)\left[\oint_{\bx^\varpi_{t',t}(C)} \bu(\br,t)\bdot d\br
\right],
     \,\,\,\,\,\,\, t_0\leq t'<t\leq t_f,
\lb{anti-stoch-kelvin} \ee
with circulations at the present time given anti-causally as averages over
future values. The process of circulations in this case is a forward
martingale.
The proofs of all of the above statements follow by straightforward
modifications
of the previous arguments for the causal case.

Conservation of energy is another property of Hamiltonian systems derived
from the least-action principle, as a consequence of time-translation
invariance.
For example, consider a standard incompressible Euler fluid, with the action
functional
\be S[\bx] = \frac{1}{2}\int_{t_0}^{t_f}dt\,\int d^dr\,\, |\bu(\br,t)|^2
\lb{euler-act} \ee
Following the procedure of Lanczos \cite{Lanczos70}, one considers an
arbitrary increasing function  $\tau(t)$ on the interval $[t_0,t_f],$ with
$\tau(t_0)=t_0$
and $\tau(t_f)=t_f,$ and defines a modified flow
\be \bx_\tau(\ba,t) = \bx(\ba,\tau(t)). \lb{tau-ex} \ee
When $\dot{\bx}(\ba,t)=\bu(\bx(\ba,t),t),$ then $\bx_\tau(\ba,t)$ satisfies the
analogous equation with
\be \bu_\tau(\br,t) = \dot{\tau}(t)\bu(\br,\tau(t)).  \lb{tau-vel} \ee
In an infinitesimal form, $\tau(t)=t+\varepsilon \delta(t)+O(\epsilon^2)$ with
$\delta(t_f)=\delta(t_0)=0,$ corresponding to a time-translation by a
time-dependent
shift. The variation in the velocity resulting  from (\ref{tau-vel}) is
$\delta\bu(\br,t)= (d/dt)[\delta(t)\bu(\br,t)],$ which implies a variation of
the action
\be \delta S[\bx]= \int_{t_0}^{t_f}dt \,\dot{\delta}(t)\int d^dr\,\,
                               \frac{1}{2}|\bu(\br,t)|^2. \lb{del-euler-act}
\ee
{}From the stationarity of the action it follows that kinetic energy $E(t)
=\frac{1}{2}\int d^dr\,\,|\bu(\br,t)|^2$ is conserved.

This argument is not valid, however, for the stochastic action principle of
Section 1.
Indeed, if $\bx^\varpi(\ba,t)$ solves the stochastic equation (\ref{sde}) for
backward
Lagrangian trajectories, then the time-reparameterized flow (\ref{tau-ex})
satisfies
\be
\hat{d}_t\bx^\varpi_\tau(\ba,t)=\bu^\varpi_\tau(\bx^\varpi_\tau(\ba,t),t)dt
                            +\sqrt{2\nu
\dot{\tau}(t)}\,\,\hat{d}\widetilde{\bW}^\varpi(t),
                            \lb{tau-sde} \ee
where $\bu^\varpi_\tau(\br,t)$ is given by the analogue of (\ref{tau-vel}) and
$\widetilde{\bW}^\varpi(t)$ is a Brownian motion defined on the same
probability
space as $\bW^\varpi(t).$ This is a consequence of standard results on
time-change
in stochastic differential equations (e.g. see \cite{IkedaWatanabe81}, Ch.IV,
section 7).
We thus see that the reparameterization (\ref{tau-ex}) leads to a flow map
which
is outside the class obeying an equation of the form (\ref{sde}) and for which
the
stochastic action (\ref{action}) is formally $+\infty.$ Thus the argument
leading
to energy-conservation based on time-translation invariance of the action is no
longer valid.

One interest of the characterization of Navier-Stokes solutions via an action
principle is that it may give some hint as to the character of their
zero-viscosity
limit. It was long ago conjectured by Onsager \cite{Onsager49} that singular
solutions of the Euler equations may result from that limit, relevant to the
description of turbulent energy dissipation at high Reynolds numbers.
For recent reviews, see \cite{Eyink08,Constantin08}. The variational principle
formulated for the Navier-Stokes solutions in the present work is similar to
that
of Brenier \cite{Brenier03} for generalized Euler solutions, in which
deterministic
Lagrangian trajectories are also replaced by distributions over histories.
There is
other evidence to suggest that this may be a physical feature of the
zero-viscosity limits
of Navier-Stokes solutions, based upon recent results in a simpler problem, the
Kraichnan model of random advection by a rough velocity field that is
white-noise
in time \cite{Kraichnan68,Falkovichetal01}. Unlike the smooth-velocity case
considered
in \cite{Kunita90}, it has been shown for the case of rough velocities that the
solutions of the stochastic equations (\ref{sde}) and (\ref{anti-sde}) for
backward and
forward Lagrangian trajectories do {\it not} become deterministic in the limit
as
$\nu\rightarrow 0$ \cite{Bernardetal98,GawedzkiVergassola00,EVanden-Eijnden00,
EVanden-Eijnden01,LeJanRaimond98, LeJanRaimond02, LeJanRaimond04}.
Instead, there are unique and nontrivial probability distributions on
Lagrangian
histories in the limit, a property referred to as ``spontaneous
stochasticity.'' This
property is a direct consequence of Richardson's law  of 2-particle turbulent
diffusion
\cite{Richardson26} and thus extends very plausibly to Navier-Stokes turbulence
in the limit of large Reynolds numbers.

The results discussed above helped to motivate the conjecture in
\cite{Eyink06,Eyink07} that a martingale property of circulations
should hold for Euler solutions obtained in the limit $\nu\rightarrow 0.$
The derivation of the stochastic Kelvin theorem in the present paper based on
particle-relabelling symmetry is closely related to a similar argument in
\cite{Eyink06}
for the martingale property of circulations in generalized Euler solutions (see
section
4 there). Note, however, that it was incorrectly proposed in \cite{Eyink06}
that
circulations for such Euler solutions should be martingales forward in time,
and this conjecture was only later emended to a backward-martingale
property in \cite{Eyink07}. The present work shows that the backward-martingale
property is indeed the natural one, which could be expected to hold for
dissipative
Euler solutions obtained as the zero-viscosity limit of Navier-Solutions
solutions.

\section{Final Remarks}

The stochastic Lagrangian representation of Constantin and Iyer
\cite{ConstantinIyer08}
and our closely related variational formulation should clearly extend to a wide
class
of Hamiltonian fluid-mechanical models with added Laplacian dissipation.
In a forthcoming paper \cite{EyinkNeto09} we prove the analogous results for
several
non-ideal (resistive and viscous) plasma models,  including the two-fluid model
of
electron-ion plasmas, Hall magnetohydrodynamics (MHD), and standard MHD. 
As we shall show there, those models possess two stochastic Lagrangian 
conservation laws, one corresponding to the Alfv\'{e}n Theorem 
on conservation of magnetic flux \cite{Alfven42} 
and another corresponding to a generalized Kelvin Theorem
\cite{KuznetsovRuban00,BekensteinOron00}.  In following work
we shall apply these results to important physical problems of magnetic
reconnection and
magnetic dynamo, especially in turbulent MHD regimes.

As noted earlier, similar stochastic least-action principles have recently been
proposed
for the incompressible Navier-Stokes equations
\cite{Rapoport00,Gomes05,CiprianoCruzeiro07}.
A detailed discussion of the relation of these different variational principles
to ours,
as well as a rigorous treatment of the latter, will be the subject of future
work.
It is worth remarking that there is another variational principle for fluid
equations,
Onsager's principle of least dissipation \cite{Onsager31,OnsagerMachlup53},
which
determines the probability of molecular fluctuations away from hydrodynamic
behavior
in terms of the dissipation required to produce them.  A modern formulation is
presented
in \cite{Eyink90}, and \cite{QuastelYau98} gives a rigorous derivation of
Onsager's
principle for incompressible Navier-Stokes in a microscopic lattice-gas model.
It would be interesting to know if any relation exists between the least-action
and
least-dissipation principles.

\bigskip

\noindent {\bf Acknowledgements.} This work was partially supported by
NSF grant AST-0428325 at Johns Hopkins University. We acknowledge
the warm hospitality of the Isaac Newton Institute for Mathematical Sciences
during the programme on ``The Nature of High Reynolds Number Turbulence'',
when this paper was completed.

\end{document}